%% file: main.tex
\documentclass[sigconf, nonacm]{acmart}
\usepackage{amsfonts}
\usepackage{multirow}
\usepackage{pifont}
\usepackage[ruled,linesnumbered]{algorithm2e}
\usepackage{subcaption}
\usepackage{graphicx}
\usepackage{xcolor}

\newcommand\vldbdoi{XX.XX/XXX.XX}
\newcommand\vldbpages{XXX-XXX}
\newcommand\vldbvolume{14}
\newcommand\vldbissue{1}
\newcommand\vldbyear{2020}
\newcommand\vldbauthors{\authors}
\newcommand\vldbtitle{\shorttitle} 
\newcommand\vldbavailabilityurl{URL_TO_YOUR_ARTIFACTS}
\newcommand\vldbpagestyle{plain} 

\usepackage{caption}

\SetKw{KwStep}{step}
\SetKw{KwRet}{return}

\begin{document}

\SetKwComment{Comment}{/* }{ */}

\title{CS-PQ: Cache-Friendly SIMD Product Quantization for Large-Scale ANNS Index Construction}
 
% \title{Accelerating Product Quantization Encoding via Cache-Friendly Parallelism for ANNS}

%\begin{comment}

%%
%% The "author" command and its associated commands are used to define the authors and their affiliations.
\author{Yutang MA}
\affiliation{%
  \institution{The Chinese University of Hong Kong}
  %\streetaddress{P.O. Box 1212}
  %\state{Ireland}
}
\email{1155173882@link.cuhk.edu.hk}

\author{Kecheng HUANG}
%\orcid{0000-0002-1825-0097}
\affiliation{%
  \institution{Beijing Institute of Technology, Zhuhai}
  %\streetaddress{1 Th{\o}rv{\"a}ld Circle}
  %\city{Hekla}
  %\country{Iceland}
}
\email{huangkecheng@bitzh.edu.cn}

\author{Xikun JIANG}
%\orcid{0000-0001-5109-3700}
\affiliation{%
  \institution{The Chinese University of Hong Kong}
  %\city{Rocquencourt}
  %\country{France}
}
\email{xkjiang24@cse.cuhk.edu.hk}

\author{Meiling WANG}
\affiliation{%
  \institution{Huawei Technologies Co., Ltd.}
  %\city{T\"ubingen}
  %\country{Germany}
}
\email{wangmeiling17@huawei.com}
%\email{myprivate@email.com}
%\email{second@affiliation.mail}

\author{Xin YAO}
\affiliation{%
  \institution{Huawei Technologies Co., Ltd.}
  %\city{Shanghai}
  %\country{China}
}
\email{yao.xin1@huawei.com}

\author{Renhai CHEN}
\affiliation{%
  \institution{Huawei Technologies Co., Ltd.}
  %\city{Duckburg}
  %\country{Calisota}
}
\email{chenrenhai@huawei.com}

\author{Gong ZHANG}
\affiliation{%
  \institution{Huawei Technologies Co., Ltd.}
  %\city{Duckburg}
  %\country{Calisota}
}
\email{nicholas.zhang@huawei.com}

\author{Zili SHAO}
\affiliation{%
  \institution{The Chinese University of Hong Kong}
  %\city{Duckburg}
  %\country{Calisota}
}
\email{shao@cse.cuhk.edu.hk}
%\affiliation{%
  %\institution{Donald's Second Affiliation}
  %\city{City}
  %\country{country}
%}
%\email{donald@swa.edu}
%\end{comment}
\begin{abstract}
\input{abs}
\end{abstract}

\maketitle

\pagestyle{\vldbpagestyle}
\begingroup\small\noindent\raggedright\textbf{PVLDB Reference Format:}\\
\vldbauthors. \vldbtitle. PVLDB, \vldbvolume(\vldbissue): \vldbpages, \vldbyear.\\
\href{https://doi.org/\vldbdoi}{doi:\vldbdoi}
\endgroup
\begingroup
\renewcommand\thefootnote{}\footnote{\noindent
This work is licensed under the Creative Commons BY-NC-ND 4.0 International License. Visit \url{https://creativecommons.org/licenses/by-nc-nd/4.0/} to view a copy of this license. For any use beyond those covered by this license, obtain permission by emailing \href{mailto:info@vldb.org}{info@vldb.org}. Copyright is held by the owner/author(s). Publication rights licensed to the VLDB Endowment. \\
\raggedright Proceedings of the VLDB Endowment, Vol. \vldbvolume, No. \vldbissue\ %
ISSN 2150-8097. \\
\href{https://doi.org/\vldbdoi}{doi:\vldbdoi} \\
}\addtocounter{footnote}{-1}\endgroup

\ifdefempty{\vldbavailabilityurl}{}{
\vspace{.3cm}
\begingroup\small\noindent\raggedright\textbf{PVLDB Artifact Availability:}\\
The source code, data, and/or other artifacts have been made available at \url{https://github.com/codinghardwear/CS-PQ}.
\endgroup
}

\section{Introduction}
\label{intro}
\input{intro/intro_v5}

% \section{Background and Motivation}
\vspace{-0.3cm}
\section{Background}
\input{bg/bg_v7}

\vspace{-0.25cm}
\section{Motivation}
\input{mot/mot_v4}

\vspace{-0.3cm}
\section{Design}
\input{tech/tech_overview_v2}

\input{tech/tech_simd_v4}
\input{tech/tech_cache_v4}
\input{tech/tech_formula_v6}

\vspace{-0.2cm}
\section{Evaluation}
\label{exp}
\input{exp}

\vspace{-0.3cm}
\section{Related Work}
\input{rw}

\vspace{-0.3cm}
\section{Conclusion}
\input{conc}

%\begin{acks}
% This work was supported by the [...] Research Fund of [...] (Number [...]). Additional funding was provided by [...] and [...]. We also thank [...] for contributing [...].
%\end{acks}

%\clearpage

\bibliographystyle{ACM-Reference-Format}
\bibliography{ref}

\end{document}

%% file: abs.tex
Product Quantization (PQ) construction is deeply integrated into vector index construction for Approximate Nearest Neighbor Search (ANNS). The rapid growth in vector dimensionality and volume has significantly increased the computational cost of PQ. 
Existing GPU-based PQ accelerations are ill-suited for PQ construction due to its “one-to-one” execution pattern (one compute, one data load, i.e., data transfer overhead dominates). Although CPU-based solutions are prevalent, they are essentially general-purpose designs that fail to capture the intrinsic characteristics of PQ construction.
In this paper, we propose CS-PQ, a Cache-friendly, SIMD-optimized PQ framework based on modern CPUs. CS-PQ introduces a vector-oriented SIMD paradigm that decouples quantization granularity from SIMD width by vectorizing across PQ centroids rather than subvector dimensions. It further restructures the execution pipeline to improve cache locality and reformulates PQ computation to eliminate redundant operations while preserving correctness. Experiments on large-scale datasets show that CS-PQ achieves up to $10.7\times$ speedup over state-of-the-art CPU-based PQ implementations without sacrificing ANNS accuracy.

%% file: intro/intro_v5.tex
Approximate nearest neighbor search (ANNS)~\cite{anns-0,anns-1,anns-2,ann-3} has become a core primitive in modern data management systems~\cite{manu,Milvus} and is widely used in applications such as vector databases~\cite{starling,diskann,hnsw,nsw-1}, information retrieval~\cite{ir-1,ir-2,ir-3}, recommendation systems~\cite{rs-0,rs-1,rs-2}, and large-scale machine learning pipelines~\cite{llm-1,llm-2,llm-3,llm-4}. In modern ANNS systems, {\bf Product Quantization (PQ)} is deeply integrated into index construction to support vector searches towards large-scale vector data~\cite{pq-1,pq-2,pq-3,pq-4}.
In recent years, advances in representation learning have led to a steady increase in vector dimensionality, with embeddings often reaching hundreds or even thousands of dimensions~\cite{scale-1,scale-2,scale-3}. At the same time, the number of vectors managed by ANNS systems continues to grow rapidly, with existing practices already operating at the scale of hundreds of millions or billions of vectors~\cite{diskann,starling}. As a result, the computational cost of PQ during index construction has increased substantially.
In addition, many real-world systems must handle continuous or frequent data insertion due to streaming inputs, model retraining, or dataset updates. Under these workloads, PQ is no longer performed only once during offline preprocessing, but is repeatedly executed during dynamic insertion, making its efficiency an important factor in large-scale ANNS index construction.

Intuitively, leveraging GPUs is a natural choice to enhance the computational efficiency of PQ construction~\cite{ann-3,gpu-1,gpu-2}. However, prior research has shown that while GPUs can accelerate PQ-based distance computations during query processing, they do not deliver the same advantages during index construction.
Specifically, GPUs excel in the ``many-to-one'' computation pattern, where dense computations are performed with minimized data transfers and kernel invocations. This aligns well with the query processing phase, where a batch of PQ-encoded vectors is loaded onto the GPU for repeated distance calculations, returning only the top candidate to the host for subsequent traversal.
In contrast, PQ applied during index construction follows a different ``one-to-one'' pattern. 
PQ is performed once per vector~\cite{diskann,starling}, meaning that for each PQ operation, the GPU receives a single high-dimensional vector, performs a one-time computation, and returns one encoded (compressed) vector to the host. Consequently, the overhead associated with transferring these high-dimensional vectors to GPU memory and then returning the resulting encoded representations to host memory becomes the dominant factor.

\begin{figure}[t]
\centering
\includegraphics[width=0.9\columnwidth]{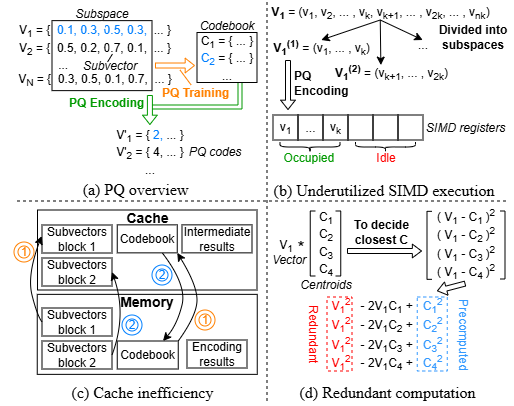}
\caption{
An Overview of PQ execution and its inefficiencies.
}
\vspace{-0.2cm}
\label{fig:intro}
\end{figure}

Recognizing the data transfer inefficiencies of GPU-based acceleration, contemporary ANNS systems increasingly favor CPU-based PQ during index construction. The objective of PQ is to reduce vector dimensions from $d$ to $m$ ($m \ll d$) by dividing the original vector into $m$ subvectors, with each subvector retaining one dimension (i.e., a PQ code). To achieve this, CPU-based PQ treats subvectors with identical dimensionality as a subspace, and encodes the subvectors into PQ codes based on a per-subspace codebook. To generate the codebook, the K-means algorithm is widely adopted, which clusters subvectors and produces centroids (e.g., $\mathbf{c}_1$). To generate the PQ code for each subvector, distance calculations are performed to find the closest centroid from the codebook (e.g., assigning PQ code 2 to compressed vector $\mathbf{v}_1'$'s subvector based on centroid $\mathbf{c}_2$). This procedure repeats until the entire vector is encoded.

\vspace{-0.4cm}
\subsection{Critical Issues}

Although opening a door for CPU-based PQ, this execution model is essentially a general-purpose approach, not tailored for workload characteristics of PQ on large-scale, high-dimensional datasets. Simply adopting this model still induces ultra-long time consumption for PQ.
We analyze three critical issues as follows.

{\noindent \bf Issue $\sharp$1: Dilemma between PQ efficiency and generation quality.}
The subvector-oriented execution model poses an unexpected constraint on computation efficiency.
In Figure~\ref{fig:intro}(a), PQ construction, including both codebook generation and PQ code generation, is performed by feeding each subvector into SIMD registers for parallel distance calculation. As each subvector involves $d/m$ dimensions, the SIMD is leveraged in a single-instruction (i.e., distance calculation), multiple-data (i.e., $d/m$ dimensions) manner. 
This execution model encounters a tough dilemma between PQ efficiency and generation quality.
Specifically, existing SIMD, represented by AVX-512, provides 512-bit width of registers for parallel processing~\cite{avx512}. Ideally, the perfect utilization of SIMD is to set $d/m$ equal to this width. However, although increasing $d/m$ to match this width contributes to better SIMD efficiency, it results in poor accuracy for future ANNS tasks as fewer PQ codes are produced after PQ compression (Figure~\ref{fig:intro}(b)). Preliminary experiments indicate that setting $d/m \le 8$ provides competitive ANNS accuracy while it is still far less than the SIMD register width provided by modern CPUs. 
Essentially, the mismatch between compression units ($d/m$) and compute lanes (width) introduces an unwanted dependency between computational efficiency and quantization quality.

{\noindent \bf Issue $\sharp$2: Asymmetric behaviors between subspace and codebook.}
This subspace-oriented execution model not only affects the computation efficiency, but also introduces redundant intermediate results that are repeatedly moved between CPU cache and main memory, which further degrades the PQ efficiency.
As shown in Figure~\ref{fig:intro}(c), taking PQ code generation as an example, to better utilize the SIMD parallelism, contemporary SIMD optimizations, represented by FAISS CPU implementation~\cite{scale-1} and Intel MKL-based vectorized distance kernels~\cite{mkl}, arrange all subvectors residing in the subspace into a matrix and then perform distance calculations between the matrix and the codebook. Although this model is proven efficient in various domains such as dense linear algebra~\cite{simd} and batched nearest neighbor search~\cite{gpu-1,gpu-2}, it fails to grasp the workload characteristics of PQ construction. The matrix processing in PQ construction is inherently asymmetric: the subspace matrix is huge since billions of vectors may be processed at a time, while the codebook is relatively small (e.g., only 256 centroids involved). Moreover, the codebook is frequently reused during computation as each subvector needs to compare with all centroids in the codebook. Intermediate results generated by this matrix processing can easily pollute the CPU cache and trigger frequent eviction and reloading for the codebook. These long-latency codebook movements result in SIMD waits, further prolonging the PQ construction time.

{\noindent \bf Issue $\sharp$3: Neglecting the ranking nature of PQ construction.}
By further investigating these intermediate results, we observe that a significant portion is redundant and contributes nothing to PQ results. Unfortunately, these redundant results and their associated computations remain largely neglected.
PQ is inherently ranking-oriented, but this feature is not favored by common SIMD optimizations.
The best match (i.e., relative distance) is sufficient for both codebook generation and PQ code generation, while absolute distance calculations, with additional overhead, are performed in common practice. Specifically, state-of-the-art (SOTA) SIMD optimizations treat this best match demand as an absolute distance calculation problem, explicitly computing complete distance expressions for each subvector against all centroids~\cite{scale-1}. In Figure~\ref{fig:intro}(d), to determine the best match (i.e., closest) centroid for subvector $\mathbf{v}_1$, a common practice is to perform absolute distance calculations for $\mathbf{v}_1$ and all centroids. However, during these calculations, $\|\mathbf{v}_1\|^2$ is explicitly redundant and would not affect the final choice. Moreover, $\|\mathbf{c}_1\|^2, \|\mathbf{c}_2\|^2, \dots$ can be precomputed and reused for not only the current subvector but also other subvectors.
These extra arithmetic operations do not change the final selection but increase PQ construction cost in terms of cache pressure and redundant computation.

\vspace{-1.2cm}
\subsection{Our Solution}
In this paper, we revisit the design of CPU-based PQ for large-scale ANNS index construction and address the aforementioned challenges through a system-level redesign. To this end, we propose {\bf CS-PQ}, a {\bf \underline{C}}ache-friendly and {\bf \underline{S}}IMD-optimized {\bf \underline{P}}roduct {\bf \underline{Q}}uantization framework that decouples the subspace-oriented PQ model from SIMD execution constraints, mitigates cache pollution and redundant data movement, and thus improves end-to-end PQ construction performance.
Rather than adhering to the conventional matrix-oriented SIMD design, we introduce a new design paradigm termed a PQ-favored, vector-oriented SIMD (pvSIMD) computation pipeline. This paradigm better aligns with the intrinsic properties of PQ workloads, enabling effective SIMD parallelization without imposing algorithmic compromises on PQ construction.   

Furthermore, we propose a PQ-oriented computation pipeline to enhance SIMD utilization on modern CPUs. Rather than mapping SIMD lanes to subvector dimensions, we organize data and computation to enable vectorized processing across multiple centroids. Partial distance results are accumulated and compared directly in registers, avoiding horizontal reductions over low-dimensional subvectors and preventing the materialization of full distance tables in memory. This design improves effective SIMD utilization and reduces instruction overhead, which is particularly important for PQ workloads with low arithmetic intensity.

We restructure the execution organization of PQ construction to improve cache locality. By reducing the active working set during construction, the large-volume, short-lived intermediate results can be eliminated. Our vectorized processing model enables aligning PQ execution order with data reuse patterns. As a result, frequently reused centroid data can remain resident in cache across continuous PQ operations, while input vectors are isolated and streamed through the CPU cache. This cache-friendly execution can significantly reduce memory traffic and mitigate cache eviction effects during bulk PQ processing.
We also reformulate the computational paradigm of PQ construction to match its ranking-oriented nature. This reformulation mitigates the arithmetic operations performed per subvector while guaranteeing that the resulting encoding decisions are identical to those produced by the original formulation.

We have developed a full-featured prototype to evaluate the effectiveness of CS-PQ, and conduct comprehensive experiments on multiple large-scale datasets spanning different vector dimensionalities. Experimental results show that compared to SOTA CPU-based PQ construction approaches, CS-PQ speeds up PQ construction by up to $10.7\times$ without degrading ANNS accuracy.

%% file: bg/bg_v7.tex
\begin{figure}[t]
  \centering
  \includegraphics[width=1\linewidth]{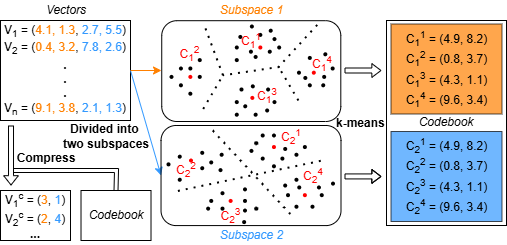}
  \caption{Illustration of the PQ workflow. }
  \vspace{-0.2cm}
  \label{fig:pq_workflow}
\end{figure}

\subsection{ANNS Index Construction and PQ}

Approximate nearest neighbor search (ANNS)~\cite{anns-1,ann-3,anns-2,anns-0} is a core primitive in modern data management systems~\cite{rag-2,rag-1,rag-3} and is widely used in vector databases~\cite{anns-0,anns-1,anns-2}, recommendation systems~\cite{rs-0,rs-2,rs-1}, information retrieval~\cite{ir-1,ir-2,ir-3}, and large-scale machine learning pipelines~\cite{llm-1,llm-2,llm-3,llm-4}. To support efficient similarity search over high-dimensional and large-scale vector data, SOTA ANNS systems rely on carefully designed index structures that balance search accuracy, memory footprint, and index construction cost~\cite{dgvm-1,survey-1,dgvm-3,dgvm-4}.

The ANNS index construction comprises three components. First, the input vector corpus is organized into a topological index structure, commonly a proximity graph, to facilitate efficient candidate set exploration during vector search processing~\cite{GGNN,nsg-1,cagra}. Second, to optimize retrieval efficiency, vector data are compressed (e.g., via product quantization or its variants) and then preserved in-memory for instant retrieval; this reduces per-vector footprint while enhancing cache locality and throughput~\cite{pq-1,pq-2,pq-3,pq-4}. Third, auxiliary metadata structures, including neighbor lists, centroid assignments, and graph connectivity information, are synthesized to accelerate distance calculation, guide index traversal, and enable dynamic insertions or deletions~\cite{survey-1,suvery-2}.

Among these subsystems, vector compression is a critical scalability enabler. As vector dimensionalities scale and corpus sizes reach hundreds of millions to billions of vectors, storing uncompressed floating-point representations becomes prohibitive, which often exceed feasible DRAM budgets by an order of magnitude. Consequently, modern ANNS systems integrate quantization within the index construction pipeline, trading marginal accuracy loss for substantial memory footprint reduction and improved memory subsystem efficiency. This design choice is not merely optional but architecturally necessary to achieve sub-linear vector search latency at wide-scale vector volumes.

{\bf Product Quantization (PQ)} is a foundational vector compression technique widely deployed in large-scale ANNS systems, which compresses each vector by quantizing its subspaces. To achieve this, PQ represents a $d$-dimensional vector by partitioning it into $m$ disjoint subvectors and independently quantizing each subvector into several bits through a dedicated codebook.

Formally, given a $d$-dimensional input vector $\mathbf{v}_i \in \mathbb{R}^{d}$, it is partitioned as $m$ independent {\tt subvectors} $\mathbf{v}_i^{(1)}, \mathbf{v}_i^{(2)}, \dots, \mathbf{v}_i^{(m)}$, and we have:
\begin{equation}
\mathbf{v}_i = [\mathbf{v}_i^{(1)}, \mathbf{v}_i^{(2)}, \dots, \mathbf{v}_i^{(m)}]
\end{equation}
where each subvector $\mathbf{v}_i^{(j)} \in \mathbb{R}^{d/m}$. For all to-be-compressed $d$-dimensional vectors $\mathbf{v}_1, \mathbf{v}_2, \dots, \mathbf{v}_i, \dots$, they can also be partitioned following the same partitioning rule. Then we define $m$ {\tt subspaces} $sp^{(1)}, sp^{(2)}, \dots, sp^{(m)}$ where each $sp^{(j)}$ represents all subvectors $\mathbf{v}_1^{(j)}, \mathbf{v}_2^{(j)}, \dots$ residing in the same subspace. For each subspace $sp^{(j)}$, a dedicated {\tt codebook} $\mathcal{C}^{(j)}$ is learned using clustering algorithms such as $k$-means.
\begin{equation}
\mathcal{C}^{(j)} = \{\mathbf{c}_1^{(j)}, \mathbf{c}_2^{(j)}, \dots, \mathbf{c}_k^{(j)}\}
\end{equation}
In this equation, each codebook $\mathcal{C}^{(j)}$ contains $k$ codes, corresponding to $k$ centroids obtained by clustering subvectors within this subspace into $k$ groups. For $\mathbf{v}_i^{(j)}$, which is originally a $d/m$-dimensional subvector, it can be represented by its closest centroid inside $\mathcal{C}^{(j)}$. This processing is termed encoding.
Specifically, during encoding, each subvector $\mathbf{v}_i^{(j)}$ is assigned to its nearest centroid based on distance calculations. We use the centroid index $z_i^{(j)}$ to represent the centroid in the codebook $\mathcal{C}^{(j)}$ we select to represent $\mathbf{v}_i^{(j)}$.
\begin{equation}
z_i^{(j)} = \arg\min_{\ell \in [1,k]} 
\left\| \mathbf{v}_i^{(j)} - \mathbf{c}_\ell^{(j)} \right\|^2 .
\label{eq:pq-assign}
\end{equation}

The same encoding processing can be performed for all subspaces based on their dedicated codebooks. The final PQ representation of $\mathbf{v}_i$ is the concatenation of centroid indices.
\begin{equation}
\texttt{PQ}(\mathbf{v}_i) = \left( z_i^{(1)}, z_i^{(2)}, \dots, z_i^{(m)} \right).
\end{equation}

During PQ construction, Equation~\ref{eq:pq-assign} is applied independently to every vector in the dataset. Since PQ operates over all vectors and all $m$ subspaces, the total computational cost scales linearly with both dataset size and vector dimensionality $d$, making PQ construction a dominant factor in large-scale ANNS index construction.

\vspace{-0.2cm}
\subsection{GPU Acceleration and CPU-Based PQ}

GPU acceleration has been extensively investigated to optimize PQ construction, yet its efficacy is highly phase-dependent~\cite{gpu-2,gpu-1}. Specifically, during vector search processing, GPUs deliver substantial throughput gains: as codebooks and to-be-searched vector codes can be preserved in GPU memory, these asymmetric distance computations across large query batches exhibit high arithmetic intensity and data reuse, enabling efficient utilization of GPU parallelism and memory bandwidth. Because each query code is decomposed into subspace codes that repeatedly access the same small codebook centroids, multiple distance lookups can reuse the resident codebook in parallel across thousands of threads, amortizing memory access costs.
In contrast, PQ during index construction presents a fundamentally distinct computational regime. Each high-dimensional vector undergoes a one-time quantization with minimal cross-vector data reuse, resulting in a {\it memory-bound, latency-sensitive} workload. During PQ construction, every input vector must be sequentially partitioned into subvectors and compared against its corresponding subspace codebook only once, with little opportunity to reuse loaded vector data across different inputs, thereby limiting arithmetic intensity and cache locality. Critically, the uncompressed input vectors typically reside in host memory, necessitating per-vector PCIe transfers to the device. For billion-scale datasets, the cumulative overhead of host–device data movement, coupled with GPU kernel launch latency and limited GPU memory occupancy due to irregular memory access patterns, often dominates the PQ construction time. Consequently, the benefits that make GPUs effective at query time fail to materialize during construction, rendering GPU offloading suboptimal.

\begin{figure*}[t]
  \centering
  \includegraphics[width=1\linewidth]{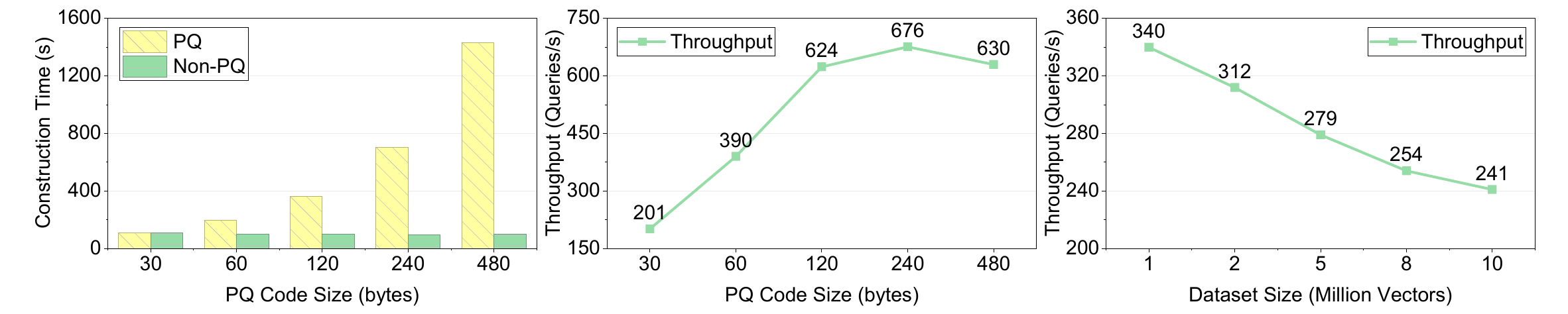}
  \vspace{-0.5cm}
    \caption{Experimental results related to the sensitive coupling of search quality and PQ overheads.}
    \vspace{-0.5cm}
  \label{fig:mot_exp}
\end{figure*}

Consequently, most contemporary ANNS systems rely on CPU-based PQ during index construction. On CPUs, PQ construction is commonly organized as a subspace-oriented, matrix-style distance computation process. Specifically, for each subspace $sp^{(j)}$, distances between a batch of subvectors ($\mathbf{v}_1^{(j)}, \mathbf{v}_2^{(j)}$, \dots) and all centroids in $\mathcal{C}^{(j)}$ are computed in a matrix-like formulation. This design resembles dense linear algebra operations and enables the reuse of SIMD-based CPU optimizations developed for general-purpose matrix workloads~\cite{simd}.
However, although this matrix-style abstraction provides a uniform and predictable execution model, it is inherently general-purpose and not specifically tailored to the workload characteristics of PQ construction over high-dimensional, large-scale vector datasets. In particular, it assumes symmetric treatment of subspaces and codebooks, materializes intermediate results in matrix form, and relies on dimension-wise SIMD parallelism over the $d/m$ dimensions.

As introduced in Section~\ref{intro}, these critical issues fundamentally constrain the performance and scalability of CPU-based PQ. In the subsequent section, we conduct a rigorous micro-architectural analysis of CPU-based PQ designs, identifying three critical inefficiencies that hinder scalability: (i) suboptimal SIMD instruction utilization due to misaligned memory access patterns; (ii) excessive cache pollution induced by non-contiguous traversal of high-dimensional vectors across subspaces; and (iii) redundant arithmetic operations arising from unoptimized centroid assignment logic.

%% file: mot/mot_v4.tex
\subsection{PQ Construction Shifts the Cost Balance}

Modern ANNS systems increasingly operate under compression-sensitive regimes driven by the interaction between dataset scale and accuracy requirements.
As data scale grows, the quality of distance approximation provided by Product Quantization becomes a dominant factor in search efficiency.
Using larger PQ code sizes preserves finer-grained information of the original vectors, which directly reduces the search effort required to achieve a target recall.
As a result, large-scale workloads naturally favor high-precision PQ configurations, corresponding to larger PQ code sizes.

This trend is reflected in practical performance measurements. We perform a preliminary evaluation to exhibit the sensitive coupling of search quality and PQ overheads. The detailed experimental setups can be found in Section~\ref{exp}.
As illustrated in Figure~\ref{fig:mot_exp}, we first profile index construction time across increasing PQ code size (i.e., total bits per vector, $m \cdot b$ where $b$ denotes the bit size per subspace, more centroids per subspace necessitate a larger $b$). 
Two clear trends emerge: First, PQ construction time increases significantly with code size due to the growth in per-subspace distance computations. Specifically, increasing $m$ increases the number of subspaces, while increasing $b$ expands the centroid cardinality ($2^b$), both of which enlarge the total number of distance computations. Second, beyond a critical code size threshold (approximately 64–96 bits per vector in our evaluation), PQ construction transitions from a secondary overhead to the dominant factor, constituting over 70\% of total construction time. This underscores PQ construction as the primary scalability bottleneck in large-scale index construction pipelines and motivates the need for architecture-aware acceleration of the quantization kernel.

We also evaluate vector search throughput at a fixed accuracy target (Recall@10 = 0.95) across varying PQ code sizes. As depicted, throughput increases substantially with code size up to a moderate configuration (e.g., 240 bytes): higher-bit representations reduce quantization error, thereby decreasing the false positive rate during candidate retrieval. This, in turn, minimizes unnecessary candidate expansion and exact distance re-ranking overhead (i.e., two dominant costs in graph-based ANNS pipelines). Consequently, these reductions directly translate to higher vector search throughput.
Furthermore, scaling the corpus size from 1M to 10M vectors induces a pronounced throughput degradation. Critically, this degradation is most severe under aggressive quantization (small code sizes), where elevated false positive rates compound memory access costs. Investing in higher-fidelity PQ representations yields desired gains in search efficiency at scale.

These results demonstrate that PQ code size is tightly coupled with the recall-efficiency trade-off. Large-scale ANNS deployments simultaneously face increasing vector dimensionality and rapidly growing data volumes.
Under such scenarios, tuning search-time parameters alone is often insufficient to compensate for short PQ codes without incurring prohibitive latency or recall degradation.
Consequently, practical systems are increasingly pushed toward operating vectors with larger PQ code sizes in order to sustain high recall and throughput at scale.

\vspace{-0.2cm}
\subsection{Limitations of Existing PQ Construction}
\label{sec:motivation-limitations}

\begin{figure}[t]
  \centering
  \includegraphics[width=1\linewidth]{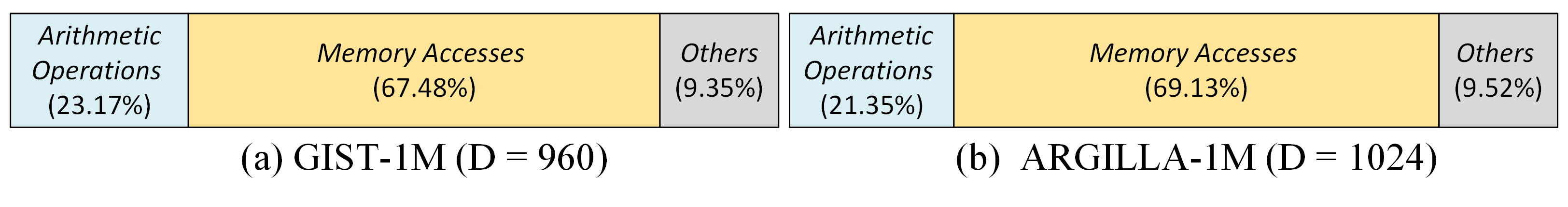}
  \vspace{-0.6cm}
    \caption{Execution breakdown of baseline PQ construction. }
    \vspace{-0.3cm}
  \label{fig:mot_memory}
\end{figure}

Figure~\ref{fig:mot_memory} presents the execution breakdown of CPU-based PQ construction time~\footnote{To distinguish between the distinct use cases of PQ, we use the term PQ construction in this paper referring to PQ during index construction.}.
The construction cost is decomposed into memory accesses, arithmetic operations, and other overhead.
Across different datasets, memory-related costs account for the largest fraction of execution time, while arithmetic operations contribute a smaller portion.
These results indicate that PQ construction exhibits memory-sensitive execution behavior on modern CPUs.

From an algorithmic perspective, PQ construction evaluates distances between each subvector and a set of codebook centroids, which suggests a computation-dominated workload.
However, the profiling results show that the practical execution cost is largely driven by the memory overheads.
This discrepancy reflects a mismatch between the abstract computation structure of PQ construction and its realized execution behavior under current formulations.

{\noindent $\bullet$ \textbf{Sub-optimal SIMD efficiency.}} PQ construction exhibits intrinsically low arithmetic intensity due to its operational characteristics during index construction. Each high-dimensional vector is partitioned into low-dimensional subvectors (typically 8–16 dimensions per subspace), and each subvector undergoes a single nearest-centroid assignment against its dedicated codebook. This workflow entails minimal computation per memory access: for a subvector of dimensionality $d/m$ and a codebook of size $k$, only $\mathcal{O}(d/m \cdot k)$ floating-point operations are performed per subvector, while the entire uncompressed vector must be loaded from memory. Crucially, intermediate results, such as partial distance computations across subspaces, exhibit negligible cross-vector reuse due to the independence of quantization across vectors and subspaces.
Consequently, the ratio of computation to data movement remains substantially below the threshold required to saturate modern CPU SIMD units. Empirical profiling reveals sustained SIMD utilization below 45\% in conventional implementations~\cite{diskann}, as short inner loops over small subspaces fail to amortize instruction-level overhead and memory-bound stalls dominate execution. Under these memory-bound conditions, scaling compute resources alone (e.g., through wider vector units or deeper pipeline parallelism) yields diminishing returns.

{\noindent $\bullet$ \textbf{Poor cache locality.}} Although PQ codebooks themselves are compact (KBs to several MBs) and exhibit high reuse potential across millions of vectors, conventional PQ construction implementations suffer from suboptimal cache utilization. Specifically, these approaches materialize full intermediate distance matrices (of size $m \times k$ per vector) prior to centroid selection. This design inflates the active working set far beyond the intrinsic codebook footprint, introducing substantial pressure on the cache hierarchy.
Consequently, despite the theoretical potential for codebook residency in L2/L3 caches, the expanded memory footprint displaces frequently accessed codebook entries, increasing cache miss rates and forcing repeated DRAM accesses for centroid data. Our preliminary experiment reveals that codebook reuse distance exceeds L3 capacity under typical batch sizes, directly translating to increased memory traffic and stalled execution units. 

{\noindent $\bullet$ \textbf{Ranking objective mismatch.}}
During PQ construction, centroid assignment within each subspace reduces to a nearest-centroid search problem: for a subvector $\mathbf{v}_i^{(j)} \in \mathbb{R}^{d/m}$, the objective is to identify the centroid $\mathbf{c}_\ell^{(j)} \in \mathcal{C}^{(j)}$ minimizing distance $\|\mathbf{v}_i^{(j)} - \mathbf{c}_\ell^{(j)}\|_2$. Conventional implementations compute the full squared distance expression for every candidate centroid:
\begin{equation}
    \left \| \mathbf{v}_i^{(j)} - \mathbf{c}_\ell^{(j)} \right \|^2 
    =  
    \underset{constant}{\left \| \mathbf{v}_i^{(j)} \right \|^2} 
    - 2\left \langle \mathbf{v}_i^{(j)}, \mathbf{c}_\ell^{(j)}  \right \rangle 
    + \underset{centroid\text{-}dependent}{\left \| \mathbf{c}_\ell^{(j)} \right \|^2}
\end{equation}

\vspace{-0.3cm}
{\noindent the term} $\left \| \mathbf{v}_i^{(j)} \right \|^2$ is invariant across all centroid comparisons within a subspace and therefore does not influence the argmin decision. Nevertheless, standard formulations redundantly evaluate this term per subvector, introducing $\mathcal{O}(d/m)$ superfluous floating-point operations that contribute neither to the encoding outcome nor to numerical stability. This arithmetic redundancy inflates instruction counts by up to 40\% in typical configurations ($k=256$), directly degrading instruction-level parallelism and exacerbating the memory-bound nature of the kernel without improving accuracy. In addition, for the centroid-dependent term $\left \| \mathbf{c}_\ell^{(j)} \right \|^2$, it can be computed once and reused multiple times for all subvectors residing in this subspace. However, it is repeatedly computed in existing practices.

\textbf{Design challenges.}
Addressing these inefficiencies necessitates a holistic co-design spanning algorithmic reformulation, data layout optimization, and instruction-level scheduling, rather than isolated micro-optimizations. Specifically, we aim to orchestrate fine-grained vectorization that achieves high occupancy despite minimal per-element computation; restructure memory access patterns to enforce codebook residency within private caches; and eliminate ranking-irrelevant computations through algebraic simplification. This multi-directional optimization requirement directly motivates our PQ construction framework introduced in the following section, which jointly addresses parallelism, memory, and arithmetic bottlenecks while preserving bit-identical encoding semantics.

%% file: tech/tech_overview_v2.tex
\begin{figure}[t]
  \centering
  \includegraphics[width=1\linewidth]{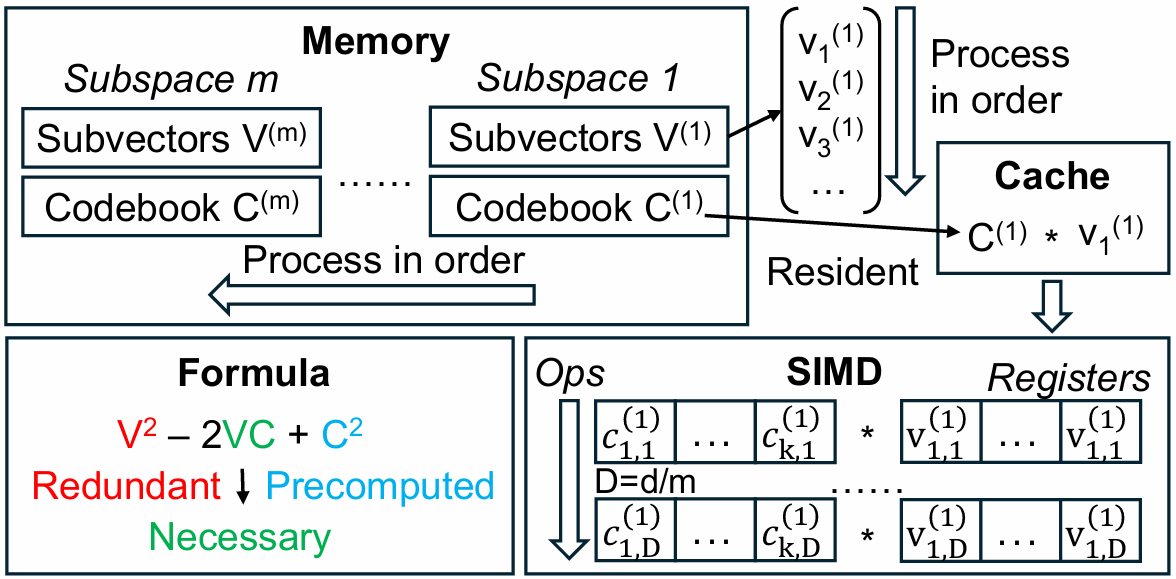}
  \caption{Design overview of CS-PQ.}
  \vspace{-0.3cm}
  \label{fig:overview}
\end{figure}

\subsection{Design Overview}

This section introduces {\bf \underline{C}}ache-friendly {\bf \underline{S}}IMD-optimized {\bf \underline{P}}roduct {\bf \underline{Q}}uantization (CS-PQ), a CPU-centric PQ construction framework to overcome the memory-bound bottlenecks of large-scale ANNS index construction. Our design co-optimizes algorithmic formulation and execution semantics to exploit three intrinsic properties of PQ construction workloads: (1) fine-grained data parallelism across low-dimensional subspaces to fully utilize SIMD-granted parallelism; (2) high reuse of compact codebooks across millions of vectors; and (3) ranking-oriented computation where only centroid ordering determines output.
To achieve this, rather than processing vectors sequentially or as a monolithic matrix, CS-PQ operates on {\tt PQ chunks}. Specifically, similar to the conventional approaches, each vector is partitioned into $m$ disjoint subvectors (PQ chunks). The key difference is rather than constructing PQ over a single matrix containing all PQ chunks, CS-PQ proceeds ''subspace-by-subspace'' with strict cache-conscious scheduling: for subspace $sp^{(k)}$, all subvectors are quantized against codebook $\mathcal{C}^{(k)}$ before advancing to $sp^{(k+1)}$. This strided execution pattern ensures that each codebook $\mathcal{C}^{(k)}$ remains resident in L2/L3 cache throughout its reuse window, eliminating redundant DRAM fetches. 
For each PQ chunk, we introduce a new design paradigm termed a PQ-favored, vector-oriented SIMD (pvSIMD) computation pipeline as follows:

\noindent\textbf{\ding{172}  Centroid-parallel SIMD computation.}
The first component performs distance computation using SIMD instructions in a centroid-parallel manner, where multiple centroids are evaluated concurrently for each subvector.
We reorganize the codebook layout to enable contiguous, vectorized access across centroids, and accumulate partial distance results directly in SIMD registers.
This design improves SIMD utilization and avoids materializing full distance tables in memory, which is critical for PQ construction workloads with low arithmetic intensity and limited data reuse.

\noindent\textbf{\ding{173} Cache-friendly execution organization.}
The second component organizes PQ construction execution in a cache-friendly, streaming manner.
Within each PQ chunk, subvectors from a block of input vectors are processed sequentially, while distance computations are performed on the fly.
This execution organization minimizes the active working set of distance accumulators and codebook accesses, allowing codebooks to remain resident in cache during bulk PQ processing and markedly reducing memory traffic.

\noindent \textbf{\ding{174} Ranking-oriented distance reformulation.}
The third component reformulates distance computation to better align with the ranking-oriented nature of PQ.
PQ construction only requires identifying the nearest centroid for each subvector, and therefore depends solely on the relative ordering of distances rather than their absolute values.
Based on this observation, we eliminate ranking-invariant terms in the distance formulation.
The reformulated distance preserves the exact centroid ranking and produces identical PQ codes as the original formulation, while reducing redundant arithmetic operations and enabling more efficient downstream execution.

%% file: tech/tech_simd_v4.tex
% \vspace{-0.2cm}
% \subsection{Centroid-Parallel SIMD Computation}
% \label{sec:stage2}

% Code identifiers: always monospaced
\newcommand{\code}[1]{\mathtt{#1}}

% Comment lines: non-monospaced + italic + gray (distinct from code)
\SetCommentSty{\itshape\color{gray}}
\newcommand{\cmt}[1]{\CommentSty{/* #1 */}\;}

% ---------- Algorithm ----------
\begin{algorithm}[!ht]
\caption{Centroid-Parallel SIMD Computation}
\label{alg:simd-accum}
\DontPrintSemicolon

\KwIn{
Subvector $\mathbf{v}_i^{(j)}\in\mathbb{R}^{d/m}$;
transposed codebook $\mathbf{C}^{(j)\top}\in\mathbb{R}^{(d/m)\times k}$;
bias terms $\mathbf{b}^{(j)}\in\mathbb{R}^{k}$ where $\mathbf{b}^{(j)}_{\ell}=\frac{1}{2}\lVert \mathbf{c}^{(j)}_{\ell}\rVert^2$;
SIMD width $w$.
}
\KwOut{PQ code index $z_i^{(j)}\in[1,k]$.}

\BlankLine
% \cmt{initialize global minimum}
$ \textit{best\_score} \gets +\infty $\;
$ \textit{best\_idx} \gets 1 $\;

\BlankLine
% \cmt{process centroids in blocks of SIMD width}
\For{$\ell \gets 1$ \KwTo $k$ \KwStep $w$}{

  % \cmt{initialize register-resident accumulator}
  $ \textit{acc} \gets \texttt{VecZero}(w) $\;

  \BlankLine
  % \cmt{accumulate inner products $\mathbf{v}_i^{(j)}\cdot\mathbf{c}^{(j)}$}
  \For{$t \gets 1$ \KwTo $d/m$}{
    $ \textit{x} \gets \texttt{VecLoad}\!\left(\mathbf{C}^{(j)\top}\!\left[t\right]\!\left[\ell:\ell+w-1\right]\right) $\;
    $ \textit{vt} \gets \texttt{VecBroadcast}\!\left(\mathbf{v}_i^{(j)}\!\left[t\right]\right) $\;
    $ \textit{acc} \gets \texttt{FMA}\!\left(\textit{x},\textit{vt},\textit{acc}\right) $\;
  }

  \BlankLine
  % \cmt{compute scores for this SIMD block}
  $ \textit{bias} \gets \texttt{VecLoad}\!\left(\mathbf{b}^{(j)}\!\left[\ell:\ell+w-1\right]\right) $\;
  $ \textit{score} \gets \textit{bias} - \textit{acc} $\;

  \BlankLine
  % \cmt{block-level reduction and global update}
  $ (\textit{blk\_min},\textit{blk\_arg}) \gets \texttt{VecArgMin}(\textit{score}) $\;

  \If{$\textit{blk\_min} < \textit{best\_score}$ \textbf{or}
      $\left(\textit{blk\_min} = \textit{best\_score} \land \ell+\textit{blk\_arg} < \textit{best\_idx}\right)$}{
    $ \textit{best\_score} \gets \textit{blk\_min} $\;
    $ \textit{best\_idx} \gets \ell + \textit{blk\_arg} $\;
  }
}
% \BlankLine
% \cmt{return centroid index with minimum score}
\Return $ \textit{best\_idx} $\;
\end{algorithm}

\vspace{-0.2cm}
\subsection{Centroid-Parallel SIMD Computation}
\label{sec:stage2}

In this section, we describe a centroid-parallel SIMD computation scheme
that exploits this structure to improve arithmetic efficiency
and reduce per-subvector instruction overhead.

As defined in Equation~\ref{eq:pq-assign},
for each subspace $sp^{(j)}$ and each subvector 
$\mathbf{v}_i^{(j)} \in \mathbb{R}^{d/m}$,
PQ construction requires evaluating its squared Euclidean distance
to all centroids in the corresponding codebook
$\mathcal{C}^{(j)} = \{\mathbf{c}_1^{(j)}, \dots, \mathbf{c}_k^{(j)}\}$:
\[
z_i^{(j)} = 
\arg\min_{\ell \in [1,k]}
\left\|
\mathbf{v}_i^{(j)} - \mathbf{c}_\ell^{(j)}
\right\|^2 .
\]

\textbf{Centroid-parallel execution.}
Existing CPU implementations of PQ construction typically apply SIMD parallelism
across dimensions or across input vectors.
However, during index construction, each vector is encoded only once
and each subvector has low dimensionality, which limits the effectiveness
of such strategies.
Instead, our design applies SIMD parallelism across centroids.
For a fixed subvector $\mathbf{v}_i^{(j)}$, scores for multiple centroids
are evaluated concurrently using wide SIMD registers.
Centroids are processed in blocks of width $w$,
corresponding to the SIMD width of the target architecture.
For each block $[\ell, \ell+w-1]$, a SIMD register maintains the partial scores
associated with these $w$ centroids
(lines~9--15 in Algorithm~\ref{alg:simd-accum}).
Iterating over centroid blocks allows all centroids to be evaluated
with a bounded and predictable register footprint.

\textbf{SIMD-friendly codebook layout.}
To support centroid-parallel SIMD access, we reorganize the codebook
into a transposed, structure-of-arrays (SoA) layout.
For each subspace $sp^{(j)}$, the transposed codebook
$\mathbf{C}^{(j)\top} \in \mathbb{R}^{(d/m) \times k}$
stores centroid coordinates such that all $k$ centroid values
for a given dimension are contiguous in memory.
Under this layout, the values
$\mathbf{C}^{(j)\top}[t][\ell:\ell+w-1]$
can be loaded using a single SIMD load
(line~11).
Compared to the conventional array-of-structures (AoS) representation,
this layout enables contiguous access across centroids
and aligns naturally with centroid-parallel SIMD computation.

\textbf{Register-resident accumulation.}
For each centroid block, accumulation registers are initialized
to zero (line~8).
The algorithm then iterates over the $d/m$ dimensions of the subvector.
At each iteration, the scalar value $\mathbf{v}_i^{(j)}[t]$ is broadcast into a SIMD register
and multiplied with centroid values loaded from
$\mathbf{C}^{(j)\top}[t][\ell:\ell+w-1]$
(lines~11--13).
The products are accumulated directly into the score registers
using fused multiply-add (FMA) instructions
(line~13).
All partial scores remain resident in registers throughout the computation,
avoiding horizontal reductions over low-dimensional subvectors
and maintaining a uniform multiply-accumulate pattern.
After all dimensions are processed,
the registers contain the inner products for the current centroid block;
the final scores are formed by combining them with the bias terms
(lines~17--18).

\textbf{Reduction and deterministic selection.}
Once the scores for a centroid block are computed,
a SIMD reduction identifies the minimum score within the block
and its corresponding centroid index
(line~21).
This block-level minimum is compared against a running global minimum.
If the block minimum is smaller, the global minimum is updated;
ties are resolved by selecting the centroid with the smaller global index
(lines~23--26).
This rule ensures deterministic behavior and reproducible PQ codes.
By structuring the kernel as an accumulate-then-reduce pipeline,
the algorithm limits live state to SIMD registers and a small number
of scalar variables.

\textbf{Discussion.}
Centroid-parallel SIMD computation restructures PQ construction
to better match low-dimensional, one-pass workloads.
By avoiding horizontal reductions and performing accumulation
entirely in registers, the computation reduces instruction overhead
and improves SIMD utilization.
Equally importantly, this register-resident execution style
produces a compact intermediate state,
which enables higher-level execution organization
to explicitly control cache behavior,
as described in Section~\ref{sec:stage3}.
Algorithm~\ref{alg:simd-accum} summarizes the SIMD computation procedure.

%% file: tech/tech_cache_v4.tex
\vspace{-0.2cm}
\subsection{Cache-Friendly Execution Organization}
\label{sec:stage3}

The centroid-parallel SIMD computation described in
Section~\ref{sec:stage2} optimizes the innermost PQ construction kernel
by keeping accumulation and reduction entirely in registers.
However, overall PQ construction performance during index construction
is also determined by how this kernel is invoked across vectors and codebooks.
In large-scale settings, even an efficient kernel can suffer from poor
end-to-end performance if its invocation pattern induces excessive
memory traffic or cache thrashing.
In this section, we describe a cache-aware execution organization
that explicitly controls the working set and data access order
during bulk PQ processing, complementing the SIMD computation kernel.

%\vspace{0.5em}
\textbf{Working set and cache pressure.}
During PQ construction, input vectors and codebooks exhibit fundamentally
different reuse patterns.
Input vectors are accessed in a streaming fashion and are typically
consumed only once during index construction.
In contrast, each codebook and its associated bias terms are reused
across a large number of vectors.
From a cache perspective, these two data types should ideally be treated
differently: codebooks should remain resident in cache, while vectors
should flow through the memory hierarchy without displacing reusable data.
Existing matrix-style PQ construction formulations fail to preserve this separation.
By expressing PQ construction as a matrix-like distance computation,
they materialize large, short-lived distance tables whose size scales
with the product of batch size and codebook size.
Although these tables are transient, they introduce substantial
write traffic and significantly inflate the effective working set.
As a result, cache capacity is dominated by intermediate distance data
that has little or no reuse.
This cache pressure repeatedly evicts codebooks from cache,
forcing them to be reloaded across batches.
Consequently, the potential reuse of codebooks is largely negated in practice,
even when the codebooks themselves are small enough to fit within cache
capacity.

%\vspace{0.5em}
\textbf{Chunk-centric execution order.}
To avoid working set inflation and preserve codebook reuse,
our design adopts a chunk-centric execution organization.
PQ is performed in blocks of input vectors to amortize
the cost of codebook access.
Within each block, the algorithm iterates over PQ chunks
as the outer loop and processes all subvectors belonging to the same
chunk before moving to the next chunk.
For a fixed chunk, the corresponding codebook and bias terms
are accessed repeatedly while subvectors from multiple vectors
are streamed through the SIMD  kernel.
This execution order creates a well-defined reuse window for each codebook.
Once a codebook is loaded into cache, it can be reused across many vectors
within the same block before eviction pressure is introduced by accesses
to other chunks.
Importantly, this reuse window is determined by the execution order
rather than by the dataset size.
As the dataset grows, the number of vectors processed within each reuse
window increases, but the size of the resident working set remains unchanged.
This property allows cache behavior to remain stable and predictable
even when PQ is applied to hundreds of millions or billions
of vectors.

%\vspace{0.5em}
\textbf{Eliminating intermediate results.}
A critical enabler of the chunk-centric execution organization
is that centroid-parallel SIMD computation does not materialize
intermediate distance results in memory.
Distance values exist only as transient register-resident state
and are reduced on the fly to a single minimum per subvector.
At no point are full distance vectors or tables written to memory.
This design choice fundamentally changes the cache footprint of PQ construction.
Because intermediate distance data never enters the cache hierarchy,
cache capacity is no longer consumed by short-lived, non-reusable arrays.
Instead, the effective working set during construction is dominated
by the codebook, its associated bias terms, and a small amount
of scalar and register state.
By eliminating intermediate result materialization,
the execution organization ensures that cache resources are primarily
allocated to data structures with high reuse potential.

%\vspace{0.5em}
\textbf{Streaming behavior and cache residency.}
Under the proposed execution organization, input subvectors are accessed
sequentially and exhibit clear streaming behavior.
Such access patterns can be efficiently handled by hardware prefetching
mechanisms, allowing vector data to be delivered to the processor
without incurring unnecessary cache pollution.
Because intermediate results are not written to memory,
streaming accesses to input vectors do not displace resident cache lines
associated with codebooks.
Meanwhile, codebooks and bias terms remain the primary resident data
in cache and can be reused across successive PQ processing operations
within the same chunk.
This separation between resident data (codebooks and bias terms)
and streaming data (input vectors) is essential for maintaining
a compact and stable working set.
It allows PQ construction performance to scale with dataset size
without relying on explicit cache management or architecture-specific tuning.

\textbf{Discussion.}
The cache-friendly execution organization operates at a different layer of the system stack than SIMD computation. While SIMD optimizes efficiency within a single kernel, execution organization governs data flow across kernels during bulk PQ processing. By eliminating intermediate distance materialization and aligning execution order with codebook reuse, the design maintains a compact and stable working set throughout index construction. This co-design of computation and execution organization addresses the memory-sensitive nature of PQ construction and enables high, predictable performance at scale while preserving identical encoding results.

%% file: tech/tech_formula_v6.tex
\subsection{Ranking-Oriented Distance Reformulation}
\label{sec:stage1}

PQ encoding assigns each low-dimensional subvector to the nearest centroid in a corresponding codebook.
Although this operation is typically expressed as a squared-distance computation,
its objective is fundamentally ranking-oriented: only the identity of the closest centroid matters,
not the absolute distance value.
However, existing implementations evaluate full distance expressions and materialize intermediate results,
which introduces unnecessary arithmetic and memory traffic.
On modern CPUs, this mismatch between computation semantics and execution behavior
complicates SIMD vectorization and increases instruction and cache overhead.
In this section, we reformulate the distance computation to preserve exact centroid ranking
while simplifying computation semantics.
This reformulation serves as the foundation for the SIMD-friendly computation
and cache-aware execution organization introduced in later stages.

\vspace{0.5em}
\noindent\textbf{Problem setup.}
An input vector of dimensionality $D$ is partitioned into $M$ PQ chunks (subspaces),
each of dimensionality $d_s = D/M$.
For a fixed chunk $m \in \{1,\dots,M\}$, its subvector
$\mathbf{v} \in \mathbb{R}^{d_s}$
is assigned to the nearest centroid from a codebook
$\mathcal{C}^{(m)} = \{\mathbf{c}_0, \dots, \mathbf{c}_{K-1}\}$,
where $K$ is the codebook size (typically $K=256$).
The PQ code for this subvector is defined as
\begin{equation}
\label{eq:pq-argmin}
\texttt{code}(\mathbf{v})
= \arg\min_{k \in [0,K-1]} \; \|\mathbf{v}-\mathbf{c}_k\|^2 .
\end{equation}
During index construction, this operation is executed once per (vector, chunk) pair
for every vector in the dataset.
Because PQ encoding is applied to all vectors and all chunks,
its cumulative cost grows linearly with dataset size and dimensionality,
making it a dominant component of index construction time at scale.

\vspace{0.5em}
\noindent\textbf{Standard distance formulation.}
A common implementation expands the squared Euclidean distance as
\begin{equation}
\label{eq:l2-expand}
\|\mathbf{v}-\mathbf{c}_k\|^2
= \|\mathbf{v}\|^2 + \|\mathbf{c}_k\|^2 - 2\,\mathbf{v}\cdot\mathbf{c}_k .
\end{equation}
While algebraically convenient, this formulation exposes a key inefficiency.
For a fixed subvector $\mathbf{v}$, the term $\|\mathbf{v}\|^2$ is constant across all centroids
and therefore cannot affect their relative ordering.
Nevertheless, conventional implementations still compute this term
and incorporate it into every distance score.
This leads to redundant arithmetic across all $K$ centroids
and inflates the amount of intermediate state that must be carried through the computation.
In practice, these redundant operations contribute to instruction overhead,
limit effective SIMD utilization, and increase pressure on registers and caches.

\vspace{0.5em}
\noindent\textbf{Ranking-oriented reformulation.}
PQ encoding only requires identifying the centroid that minimizes the distance.
For any function $\alpha(\mathbf{v})$ that depends solely on the subvector $\mathbf{v}$
and is independent of the centroid index,
\begin{equation}
\label{eq:ranking-invariance}
\arg\min_k \|\mathbf{v}-\mathbf{c}_k\|^2
=
\arg\min_k \left(\|\mathbf{v}-\mathbf{c}_k\|^2 - \alpha(\mathbf{v})\right).
\end{equation}
By setting $\alpha(\mathbf{v})=\|\mathbf{v}\|^2$ and applying it to
Equation~\ref{eq:l2-expand}, we obtain an equivalent optimization objective:
\begin{equation}
\label{eq:score-no-v2}
\arg\min_k \left(\|\mathbf{c}_k\|^2 - 2\,\mathbf{v}\cdot\mathbf{c}_k\right).
\end{equation}
For implementation convenience and to expose a regular computation structure,
we divide by $2$ and define the transformed score
\begin{equation}
\label{eq:score-reform}
s(\mathbf{v},\mathbf{c}_k)
= \frac{1}{2}\|\mathbf{c}_k\|^2 - \mathbf{v}\cdot\mathbf{c}_k .
\end{equation}
The positive scaling factor does not affect the argmin.
This reformulation removes the subvector-dependent norm term entirely
and expresses PQ encoding as an inner-product-dominant computation,
in which each centroid contributes through a uniform multiply-accumulate pattern.
As a result, the computation becomes more regular and better aligned
with SIMD execution on modern CPUs.

\vspace{0.5em}
\noindent\textbf{Correctness.}
Let
$D_k = \|\mathbf{v}-\mathbf{c}_k\|^2$
and
$S_k = \frac{1}{2}\|\mathbf{c}_k\|^2 - \mathbf{v}\cdot\mathbf{c}_k$.
From Equation~\ref{eq:l2-expand},
\[
D_k = \|\mathbf{v}\|^2 + 2S_k .
\]
For any two centroids $i$ and $j$,
$D_i < D_j$ if and only if $S_i < S_j$.
Therefore, minimizing $S_k$ preserves the exact centroid ranking for every subvector,
and the reformulated computation produces identical PQ codes
to the baseline squared-distance formulation.
This equivalence holds for all vectors and all codebooks used during index construction.

\vspace{0.5em}
\noindent\textbf{Offline and online computation.}
Equation~\ref{eq:score-reform} naturally separates PQ encoding into two components.
The offline component is performed once per codebook, where centroid bias terms
$b_k=\frac{1}{2}\|\mathbf{c}_k\|^2$ are precomputed and stored.
These values are reused across all vectors encoded with the same codebook,
thereby amortizing their computation cost over the entire dataset.
The online component is executed per subvector.
For each subvector, inner products $\mathbf{v}\cdot\mathbf{c}_k$ are computed across all $K$ centroids,
and the minimum of $b_k-\mathbf{v}\cdot\mathbf{c}_k$ is selected.
This structure avoids computing $\|\mathbf{v}\|^2$ entirely
and reduces distance evaluation to a sequence of regular multiply-accumulate operations with minimal intermediate state.

% \vspace{0.5em}
% \noindent\textbf{Discussions.}
% Although the asymptotic complexity remains dominated by $\Theta(K d_s)$,
% the reformulation removes redundant arithmetic and simplifies the computation datapath.
% More importantly, it changes the execution shape of PQ encoding.
% Distance accumulation and comparison can be performed incrementally,
% allowing partial results to be maintained in registers
% and avoiding the materialization of full distance tables in memory.
% This property enables centroid-parallel SIMD computation in the next stage
% and reduces the effective working set during encoding.
% As a result, execution can be organized so that codebooks and their associated bias terms
% remain resident in cache across successive vectors,
% forming the basis for the cache-aware execution organization
% described in Section~\ref{sec:stage3}.

%% file: exp.tex
\subsection{Experimental Setup}
\label{sec:experimental-setup}

\textbf{System Configuration.}
All experiments are conducted on a dedicated server to ensure a controlled and reproducible evaluation environment.
The server is equipped with a 40-core (80 vCPU) Intel Xeon Gold 6230 CPU, 375~GB of DDR4 main memory, and a high-performance SSD array with approximately 1.7~TB capacity. The processor supports AVX2 and AVX-512 instruction sets, which are fully enabled in our implementation.
The system runs Ubuntu 20.04 LTS with Linux kernel 5.15 and the Ext4 file system.
Unless otherwise specified, all evaluations employ a fixed thread count of 80 to saturate available hardware parallelism, with thread affinity pinned to physical cores to minimize migration overhead. All implementations are compiled with GCC 11.4.0 using -O3 -march=native -flto -fopenmp flags to enable architecture-specific optimizations.

\begin{table}[t]
\centering
\small
\caption{Dataset specifications.}
\label{tab:datasets}
\vspace{-0.3cm}
\scalebox{0.9}{
\begin{tabular}{lcccl}
\toprule
\textbf{Dataset} & \textbf{\#Vectors} & \textbf{Dim} & \textbf{\#Queries} & \textbf{Type} \\
\midrule
SIFT100M-1024D ~\cite{sift1b}      & 100M & 1024 & 10{,}000  & float32 (synthetic) \\
ARGILLA21M ~\cite{argilla21m}   & 21M  & 1024 & 100{,}000 & float32 \\
ANTON19M ~\cite{anton19m}    & 19M  & 1024 & 100{,}000 & float32 \\
LAION100M ~\cite{laion100m}   & 100M & 768  & 100{,}000 & float32 \\
SIFT100M-768D   & 100M & 768 & 10{,}000  & float32 (synthetic) \\
SIFT100M-512D     & 100M & 512 & 10{,}000  & float32 (synthetic) \\
SSNPP100M ~\cite{ssnpp100m}   & 100M & 256  & 100{,}000 & float32 \\
\bottomrule
\end{tabular}
}
\vspace{-0.3cm}
\end{table}

\noindent\textbf{Datasets.}
We conduct evaluations on seven large-scale vector datasets with varying dimensionalities and data sources, as summarized in Table~\ref{tab:datasets}.
The \textsc{SIFT100M} datasets are synthesized from the original SIFT1B benchmark to enable controlled experiments under different dimensionalities while keeping the data distribution fixed.
This allows us to isolate the impact of vector dimensionality on PQ construction behavior.
The remaining datasets consist of real-world embeddings collected from different application domains.
Each dimension is single-precision floating-point.
No dimensionality reduction (e.g., PCA) is applied unless explicitly stated.

\noindent\textbf{PQ Parameters.}
We adopt standard 8-bit per-subspace Product Quantization (PQ8), corresponding to $K=256$ centroids per subspace.
Unless otherwise specified, we fix the subvector dimensionality to 16, resulting in a 64$\times$ compression ratio relative to the original \texttt{float32} representation.
Under this configuration, the PQ code size scales proportionally with the original dimensionality (e.g., 64 bytes for 1024-dimensional vectors and 16 bytes for 256-dimensional vectors).
This compression setting is commonly used in large-scale ANN systems and provides a balanced trade-off between memory footprint, quantization fidelity, and time consumption~\cite{diskann,pq-2}.
The default codebook size is set to $K=256$, and all distances are computed using the squared Euclidean (L2) metric~\cite{survey-1}.

\noindent\textbf{Construction Parameters.}
We use the standard DiskANN implementation~\cite{diskann} as the baseline.
%Index construction follows DiskANN's default configuration for large-scale datasets.
Specifically, we set the maximum graph out-degree to $R=64$ and the candidate pool size during construction to $L=128$.
Indices are built in a block-based manner to accommodate memory constraints when processing large datasets.
Our optimized implementation modifies only the PQ construction stage of the DiskANN construction pipeline; all other components, including graph construction, neighbor pruning, and index layout, remain unchanged to ensure a fair comparison.

\noindent\textbf{Measurement.}
We primarily report PQ construction time as well as overall index build time.
Each experiment is executed multiple times, and we report the average execution time over five runs.
A warm-up run is performed before measurement to eliminate cold-start effects.
We observe minimal variance across runs and therefore omit error bars unless otherwise noted.

\begin{figure}[t]
  \centering
  \includegraphics[width=0.48\textwidth]{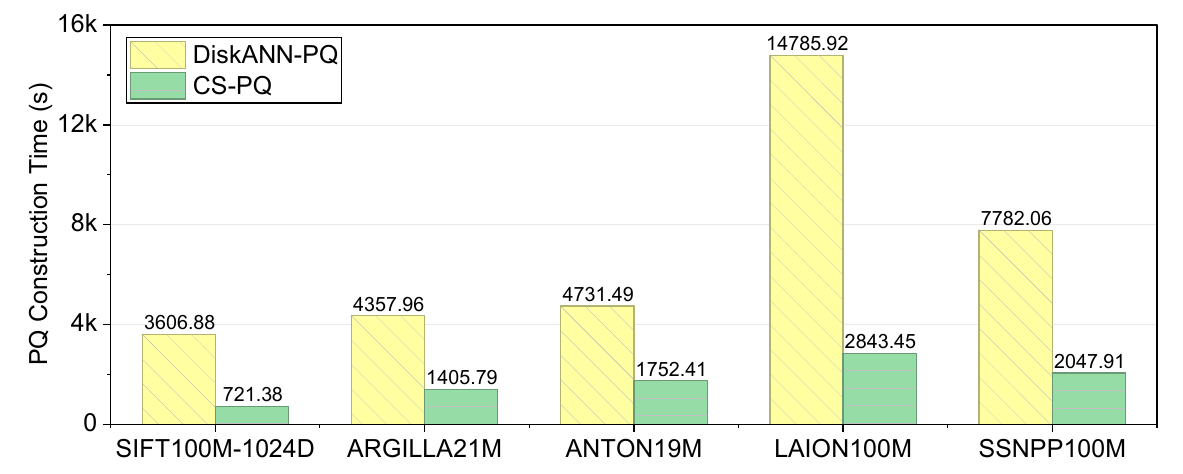}
  \vspace{-0.3cm}
  \caption{Overall comparison of PQ construction time.}
  \label{fig:exp_time}
      \vspace{-0.3cm}
\end{figure}

\begin{figure*}[t]
  \centering
  \includegraphics[width=1\textwidth]{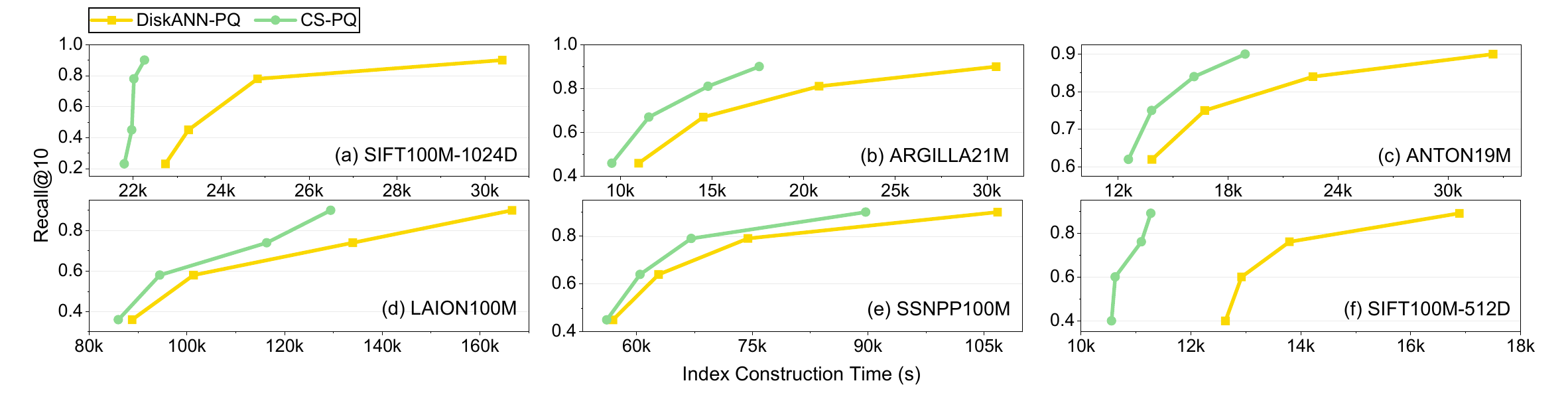}
  \vspace{-0.3cm}
  \caption{Overall comparison of index construction time at comparable Recall@10 levels.}
  \label{fig:exp_recall}
      \vspace{-0.4cm}
\end{figure*}

\begin{figure*}[t]
  \centering
  \includegraphics[width=1.02\textwidth]{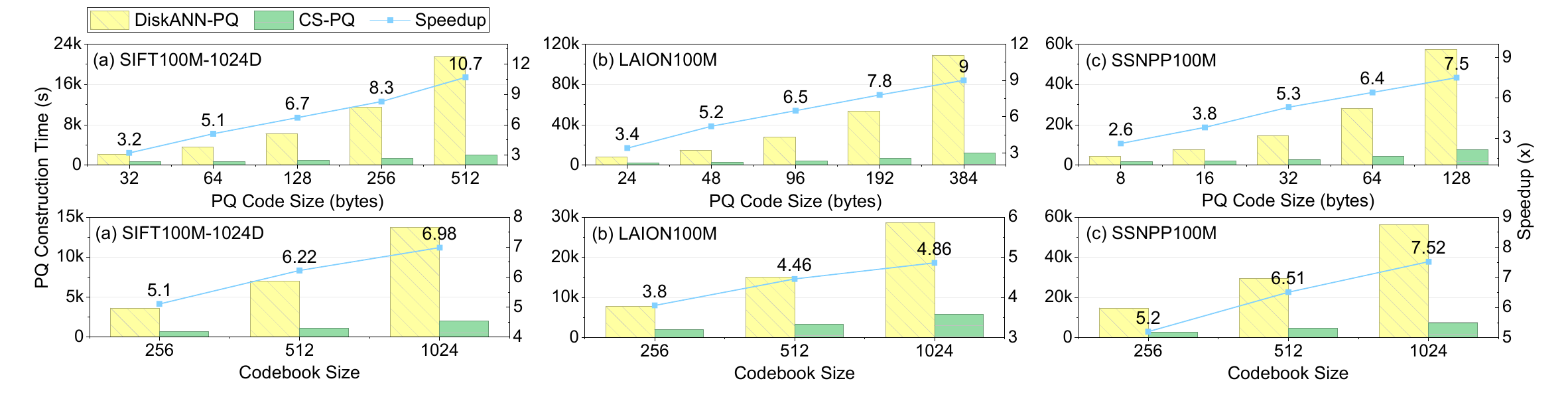}
  \caption{PQ construction time comparison under varying PQ code and codebook sizes.}
  \label{fig:exp_codesizek&codebook}
      \vspace{-0.4cm}
\end{figure*}

\vspace{-0.6cm}
\subsection{Overall Performance}
\label{sec:overall}

We first evaluate the overall impact of \textsc{CS-PQ} on index construction from two complementary perspectives: the cost of PQ construction itself and its effect on the end-to-end construction-accuracy trade-off.
Figures~\ref{fig:exp_time} and~\ref{fig:exp_recall} summarize the results from both perspectives across the evaluated datasets.

{\bf PQ construction time.}
Figure~\ref{fig:exp_time} shows the total PQ time under five distinct datasets during index construction.
Across all datasets, \textsc{CS-PQ} consistently reduces PQ construction cost, achieving speedups ranging from approximately $2.7\times$ to over $5.2\times$ compared to the baseline PQ implementation of DiskANN (\textsc{DiskANN-PQ} ).
The improvement is observed on datasets spanning a wide range of dimensionalities (from 256 to 1024) and data sources, indicating that the benefit is not specific to a particular vector distribution.
The magnitude of improvement correlates with the effective PQ workload. Higher-dimensional datasets and those using larger PQ codes (e.g., ARGILLA, ANTON, and LAION) benefit more from \textsc{CS-PQ}, as \textsc{DiskANN-PQ} performs a larger amount of ranking-invariant computation and materializes larger intermediate states.
In addition, after pvSIMD restructuring PQ construction to maintain a compact execution footprint, \textsc{CS-PQ} exhibits favorable scaling behavior as dimensionality increases.
Even for lower-dimensional datasets such as SSNPP, where PQ construction is not the sole bottleneck, \textsc{CS-PQ} still delivers noticeable reductions in PQ time.
These results reflect the efficacy of \textsc{CS-PQ} in structural improvements to the encoding pipeline rather than dataset-specific effects.

\textbf{Trade-off of construction time and search accuracy.}
Figure~\ref{fig:exp_recall} plots index construction time versus Recall@10  (i.e., recall at top-10 nearest neighbor search) across varying PQ code sizes. \textsc{CS-PQ} consistently achieves higher Recall@10 than \textsc{DiskANN-PQ} at equivalent construction latencies. At low recall rates, construction time is dominated by non-PQ phases (e.g., graph initialization and neighbor pruning), since coarse-grained PQ configurations, with lower overhead to establish, can satisfy the low recall rates requirement. As recall requirements increase, finer-grained PQ codes become necessary, causing PQ to dominate the construction pipeline. In high-recall scenarios (Recall@10 > 0.9), i.e., the operational target for most production ANNS systems, PQ acceleration directly translates to proportional end-to-end construction speedups, with \textsc{CS-PQ} reducing total build time by up to 4.1$\times$ while preserving search quality.

Accordingly, the improvement varies across datasets in a foreseeable manner.
For datasets with higher dimensionality or finer-grained PQ requirements, \textsc{CS-PQ} exhibits preferable gains in providing competitive indexing quality with substantially reduced construction time.
Overall, these results demonstrate that \textsc{CS-PQ} not only reduces the cost of PQ construction in isolation, but also reshapes the cost structure of index construction.
By lowering the dominant compression cost without affecting recall, \textsc{CS-PQ} enables faster index construction with high quality.

\begin{figure}[t]
  \centering
  \includegraphics[width=0.45\textwidth]{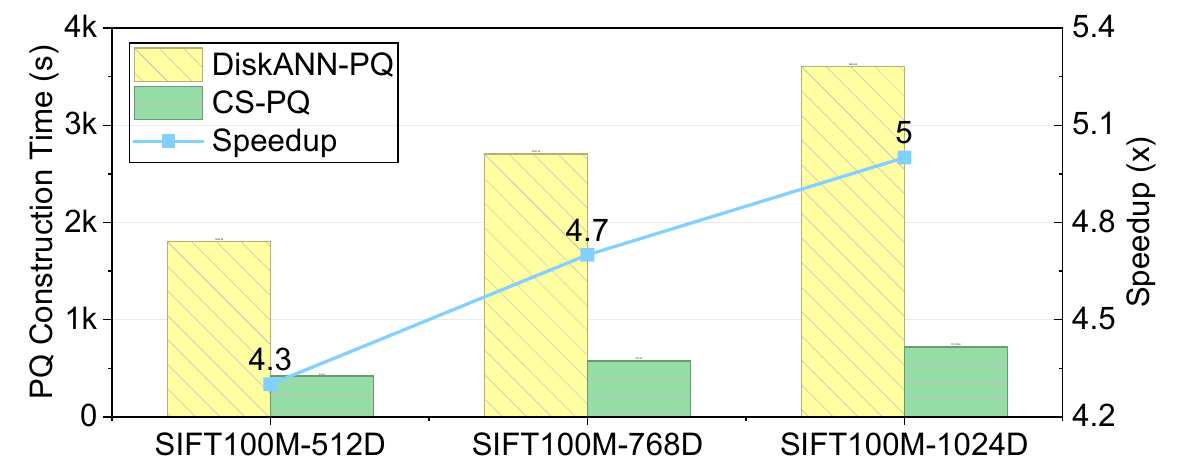}
  \caption{PQ construction time comparison under varying vector dimensionalities.}
  \label{fig:exp_dim}
      \vspace{-0.4cm}
\end{figure}

\vspace{-0.4cm}
\subsection{Microbenchmark Analysis}
\label{sec:micro_pq}

This section presents microbenchmarks that isolate PQ stages and examine how \textsc{CS-PQ} behaves under the critical parameters: PQ code size, codebook size, and vector dimensionality, respectively. In this evaluation, we use three large-scale benchmarks—SIFT100M-1024D, LAION100M, and SSNPP100M—for the code size and codebook size studies, and SIFT100M with dimensionalities of 512D, 768D, and 1024D for the dimensionality study. 
These evaluations aim to explain why \textsc{CS-PQ} delivers consistent improvements by analyzing how its benefits evolve as the intrinsic PQ workload increases.

\textbf{Effect of PQ Code Size.}
In Figure~\ref{fig:exp_codesizek&codebook}(top), we first vary PQ code size while keeping other parameters fixed. Increasing PQ code size corresponds to higher-precision representation, which is increasingly favored in large-scale ANNS systems to sustain high recall and throughput.
As shown in Figure~\ref{fig:exp_codesizek&codebook}(top), \textsc{CS-PQ} achieves progressively larger speedups over \textsc{DiskANN-PQ} as PQ code size increases across all tested ranges. The advantage grows monotonically with code size on all evaluated datasets, demonstrating that \textsc{CS-PQ} scales more effectively as the intrinsic PQ workload becomes heavier.
This is because \textsc{DiskANN-PQ} increasingly suffers from two side-effects as PQ code size increases. First, larger PQ codes amplify ranking-irrelevant computation.
The baseline formulation evaluates additional arithmetic terms whose values do not affect centroid ordering, causing instruction count to grow faster than the useful work required for encoding decisions.
Second, larger code sizes exacerbate cache pressure by increasing the volume of intermediate state and streamed data, further degrading execution efficiency.
Differently, \textsc{CS-PQ} optimizes the side-effects by (1) Ranking-oriented distance reformulation that ensures PQ precision translating directly into useful computation rather than redundant arithmetic; (2) Register-resident accumulation that avoids materializing intermediate distance results in memory.
As a result, CS-PQ scales more gracefully with PQ code size, contributing to better speedups at higher-precision scenarios.

\textbf{Effect of codebook Size.}
We then evaluate the effects of increased codebook size. In this evaluation, vector dimensionality and compression ratio remain the default setting.
Larger codebooks are commonly used to improve quantization fidelity, but they also increase the number of candidate centroids evaluated per subvector.
As shown in Figure~\ref{fig:exp_codesizek&codebook}(bottom), across all tested settings, \textsc{CS-PQ} exhibits increasing PQ time reductions as codebook size grows.
On the other hand, this trend also highlights a key limitation of existing PQ approaches: as the number of centroids increases, \textsc{DiskANN-PQ} amplifies memory traffic and cache interference due to less structured access patterns and expanded intermediate state.
Even when codebooks themselves are small enough to fit in cache, reuse is undermined by cache pollution from transient distance data.
In contrast, \textsc{CS-PQ} is explicitly designed to scale with codebook size.
Our pvSIMD execution pipeline processes multiple centroids concurrently using contiguous memory access enabled by the transposed codebook layout.
More importantly, the cache-friendly execution organization ensures that codebooks remain resident in cache while subvectors stream through the computation.
As a result, increasing codebook size can positively increase SIMD's computation density while the memory overheads are well mitigated, which allows \textsc{CS-PQ} to convert additional PQ complexity into predictable compute cost.
These results further demonstrate that \textsc{CS-PQ} does not merely optimize a fixed configuration, but remains effective as systems move toward larger codebooks to meet accuracy requirements.

\textbf{Effect of Dimensionality.}
We also study the effect of vector dimensionality across datasets with different embedding sizes, for which we assign a fixed compression ratio, i.e., PQ8 with fixed subvector dimensionality, which is a common practical setting where systems adopt a standard PQ configuration.
Figure~\ref{fig:exp_dim} shows that \textsc{CS-PQ} delivers significant speedups across all tested dimensionalities, indicating that its effectiveness is not tied to a particular dataset.
While the absolute speedup varies with dimensionality, the qualitative behavior remains stable:
\textsc{CS-PQ} reduces PQ time by 76.7\%, 78.7\%, and 80.0\% under SIFT100M-512D, SIFT100M-768D, and SIFT100M-1024D, respectively.
Higher dimensionality increases the amount of arithmetic per subvector, which improves the payoff of centroid-parallel SIMD execution.
At the same time, \textsc{DiskANN-PQ} becomes more sensitive to cache behavior as the volume of streamed data grows.
\textsc{CS-PQ} alleviates this sensitivity by maintaining a compact working set and separating streaming vector access from cache-resident codebooks.
Importantly, these experiments demonstrate that fixing compression ratio does not diminish the speedup gains of PQ optimization.

% {\noindent \bf $\bullet$ Discussion.}
% Taken together, the microbenchmarks show that \textsc{CS-PQ} systematically addresses intrinsic inefficiencies in conventional PQ construction.
% Across all these experiments, \textsc{CS-PQ} comprehensively improves PQ efficiency. The improvement correlates strongly with factors that amplify ranking-invariant computation, SIMD inefficiency, and cache pressure in baseline implementations, which precisely are the optimization directions of CS-PQ.
% These results complement the end-to-end findings in Section~\ref{sec:overall}: without relying on dataset-specific properties or relaxed accuracy constraints, the performance gains of our approach are achieved by the pvSIMD PQ execution pipeline and corresponding optimizations.

\begin{figure}[t]
  \centering
  \includegraphics[width=\linewidth]{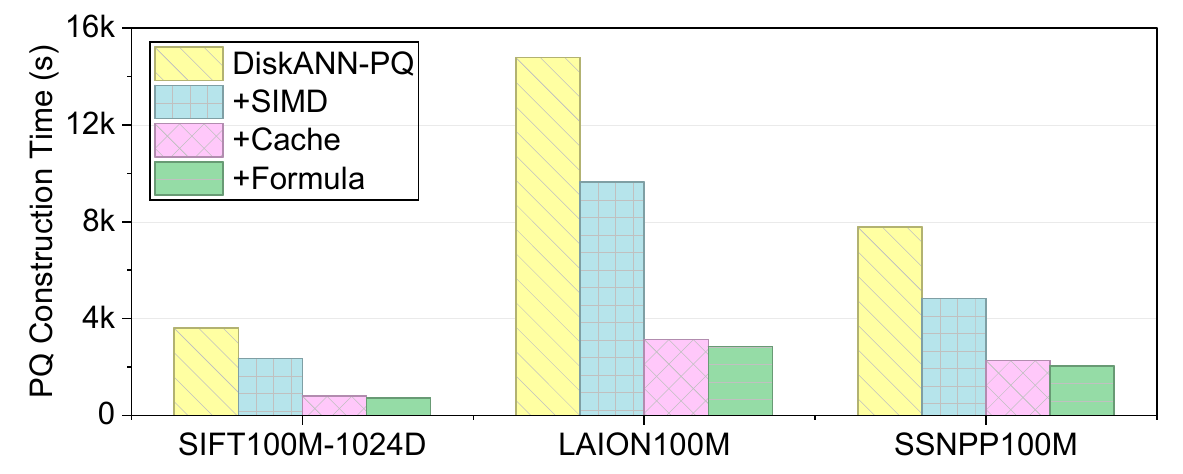}
    \caption{Impact of individual optimization components on PQ construction time.}
  \label{fig:ablation}
      \vspace{-0.2cm}
\end{figure}

\vspace{-0.5cm}
\subsection{Ablation Study}
\label{sec:ablation}

In this section, we study the effectiveness of CS-PQ's design components by incrementally enabling the key optimizations shown in Figure~\ref{fig:ablation}.
Starting from \textsc{DiskANN-PQ}, we progressively add (1) centroid-parallel SIMD execution, (2) cache-friendly execution organization, and (3) ranking-oriented distance reformulation, and then measure the PQ time consumption on three large-scale datasets under the same index configuration as introduced in Section~\ref{sec:experimental-setup}.

\textbf{Baseline} \textsc{DiskANN-PQ} incurs the highest PQ construction time across all datasets, taking approximately 3.6K seconds on SIFT100M-1024D, 14.5K seconds on LAION100M, and 7.8K seconds on SSNPP100M (Figure~\ref{fig:ablation}).
This is because the baseline suffers from both inefficient SIMD usage and memory-system inefficiencies, leaving substantial headroom for co-design across computation and cache behavior.

\textbf{+SIMD.}
Enabling centroid-parallel SIMD reduces construction time to roughly 2.4K seconds, 9.6K seconds, and 4.8K seconds on the three datasets, yielding about 1.5$\times$--1.6$\times$ speedup over \textsc{DiskANN-PQ}.
These improvements demonstrate that, once SIMD lanes are mapped to centroid blocks and score accumulation is kept register-resident, instruction overhead and suboptimal vector-lane utilization become materially
reduced.
However, the remaining time is still large especially on LAION100M, showing  that memory traffic and cache misses remain the critical factors for PQ efficiency.

\textbf{+Cache.}
Adding cache-friendly execution organization delivers the largest incremental gain, which reduces the PQ construction time to about 0.8K seconds (SIFT100M-1024D), 3.2K seconds (LAION100M), and 2.3K seconds (SSNPP100M), respectively.
Compared to the DiskANN baseline, this corresponds to $\sim$4.5$\times$, $\sim$4.5$\times$, and $\sim$3.3$\times$ speedups, respectively.
The results demonstrate that PQ construction is primarily memory-sensitive at scale: reorganizing the execution order to preserve codebook residency and shrinking the active working set directly reduces cache thrashing and DRAM traffic, which in turn translates into substantial end-to-end construction time reduction.

\textbf{+Formula (CS-PQ).}
Finally, enabling ranking-oriented distance reformulation also yields substantial PQ construction time reductions, bringing construction time down to about 0.7K seconds, 2.8K seconds, and 2.0K seconds on the three datasets.
Overall, CS-PQ achieves end-to-end speedups of approximately 5.5$\times$ (SIFT100M-1024D), 5.2$\times$ (LAION100M), and 3.9$\times$ (SSNPP100M) compared to \textsc{DiskANN-PQ}.
At this stage, the reformulation mainly removes ranking-invariant arithmetic and simplifies the score data path, which complements the pvSIMD pipeline. 

\begin{figure}[t]
  \centering
  \includegraphics[width=0.5\textwidth]{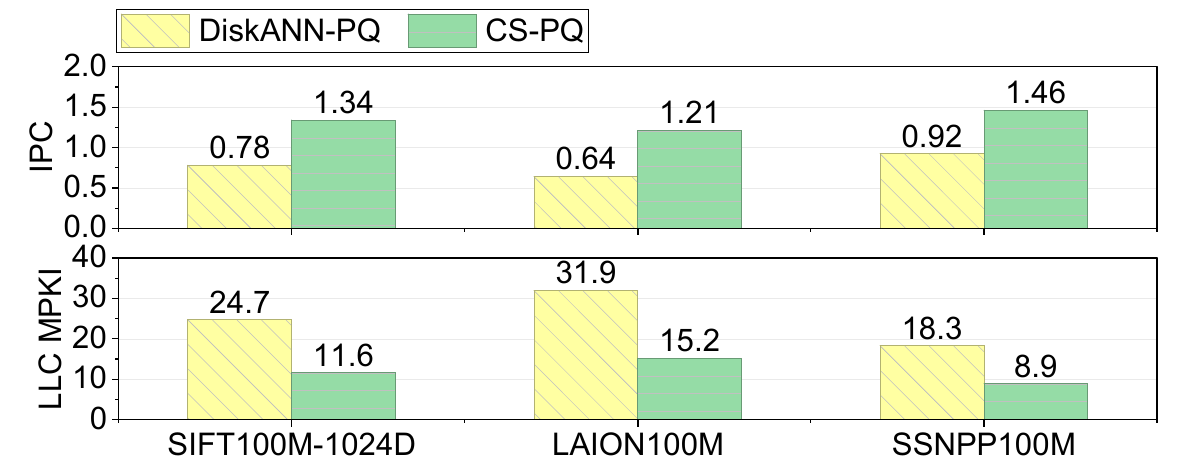}
  \caption{Microarchitectural analysis of IPC and LLC MPKI.}
  \label{fig:evidence}
      \vspace{-0.2cm}
\end{figure}

\vspace{-0.35cm}
\subsection{Microarchitectural Evidence}
\label{sec:evidence}

To further demonstrate the PQ construction time reduction, we evaluate the microarchitectural behavior of PQ construction during index construction.
Figure~\ref{fig:evidence} illustrates two hardware-level metrics, instructions per cycle (IPC) and last-level cache misses per thousand instructions (LLC MPKI), which are measured during PQ construction for both the baseline and CS-PQ.

\textbf{IPC.}
In Figure~\ref{fig:evidence}, CS-PQ consistently achieves higher IPC than the baseline across all evaluated datasets.
For the baseline, IPC remains below 1.0, indicating that execution is frequently stalled and that available compute resources are not fully utilized.
For CS-PQ, IPC increases to above 1.2, reflecting improved utilization of the processor execution pipeline.
This improvement demonstrates that CS-PQ achieves better effectiveness in SIMD utilization.
%The observed IPC increase is consistent with the execution restructuring introduced by successive ablations, which reduces stall cycles and enables steadier progress during PQ construction.

\textbf{LLC MPKI.}
Figure~\ref{fig:evidence} also shows that CS-PQ substantially reduces LLC MPKI compared with the baseline.
Across all evaluated datasets, CS-PQ decreases LLC MPKI by 51.4\% to 53.0\% compared to the baseline, which demonstrates that PQ construction under CS-PQ incurs significantly fewer last-level cache misses.
The lower miss rate indicates a more stable active working set, allowing frequently accessed codebooks to be reused more effectively before eviction, and thus reducing long-latency memory accesses.

Viewed jointly, the IPC and LLC MPKI measurements provide concrete evidence for the bottleneck shift identified in
Section~\ref{sec:ablation}.
The early-stage optimizations primarily improve instruction-level efficiency, and the pronounced reduction in LLC MPKI shows that memory behavior plays a decisive role in the final performance gains.
Fewer cache misses reduce memory stall time, enabling the processor to sustain higher IPC and make steady progress through the pvSIMD pipeline.
Consequently, improvements in cache locality and memory access efficiency are directly reflected in lower PQ construction time.
These results demonstrate that the performance gains of CS-PQ stem from coordinated improvements in both computation efficiency and cache behavior.

%% file: rw.tex
Product Quantization (PQ) is a foundational vector compression technique for large-scale approximate nearest neighbor (ANN) search. By decomposing a high-dimensional vector into multiple subspaces and independently quantizing each subvector, PQ balances memory footprint, approximation fidelity, and distance evaluation efficiency. Numerous extensions refine the basic formulation, including rotation-based variants such as Optimized PQ (OPQ)~\cite{opq}, residual and multi-stage quantization schemes~\cite{aq,cq}, and other accuracy-oriented improvements~\cite{ivfpq} that reduce quantization error or enhance search recall. These efforts primarily target representation quality or accelerate query-time distance evaluation through lookup tables and asymmetric distance computation.

Beyond model-level refinements, PQ has been deeply integrated into
modern ANN index designs~\cite{diskann,starling,pq-1,pq-4}, particularly in graph-based systems where
compressed vectors are stored in memory to reduce bandwidth pressure
during search. In such systems, PQ is often treated as a fixed
compression primitive, while optimization efforts concentrate on graph
traversal, candidate pruning, and query-time scoring. However, as
vector dimensionality and dataset scale continue to grow, the cost of
applying PQ across the entire dataset during index construction becomes
increasingly significant. In high-precision configurations required to
sustain high recall, PQ can dominate construction time, exposing
system-level inefficiencies that are orthogonal to the quantization
model itself.

Prior works on SIMD PQ acceleration and
cache-aware optimization have provided general techniques for improving
distance computation kernels~\cite{quickeradc}, including vectorized fused
multiply-accumulate pipelines~\cite{anatomy}, data layout transformations~\cite{anatomy,cache}, and
blocking strategies to enhance locality\cite{cache}. While these approaches are
effective for dense linear algebra and high-arithmetic-intensity
workloads, they cannot directly be applied to PQ construction due to its distinct characteristics.
%These general-purpose PQ acceleration
%may therefore inflate intermediate state, reduce SIMD lane utilization,
%and disrupt cache residency of codebooks.

In contrast to prior efforts that modify the PQ model or focus
primarily on query-time acceleration, our work focus on PQ construction. By aligning
distance computation with PQ’s ranking objective, vectorizing across
centroids, and structuring execution to preserve codebook residency in
cache, we systematically address the arithmetic redundancy, SIMD underutilization, and
memory-system inefficiency issues while preserving
bit-identical quantization semantics.

%% file: conc.tex
We present \textsc{CS-PQ}, a cache-friendly, SIMD-optimized PQ construction pipeline for CPU-based large-scale ANNS index construction. \textsc{CS-PQ} realigns the PQ construction pipeline with modern SIMD architecture to reduce redundant calculations, improve SIMD efficiency, and preserve codebook residency in cache during bulk PQ processing. Experiments on large-scale datasets show that \textsc{CS-PQ} consistently reduces PQ construction cost and improves the end-to-end index construction time without affecting recall.